\documentclass[aps,prb,twocolumn,showpacs,preprintnumbers,superscriptaddress]{revtex4}
\usepackage{amssymb}
\usepackage{graphicx}
\usepackage{dcolumn}
\usepackage{bm}
\usepackage{color}

\begin{document}

\preprint{PREPRINT (\today)}
\title{Quantum superconductor-insulator transition: Implications of
BKT-critical behavior}
\author{T.~Schneider}
\email{toni.schneider@swissonline.ch}
\affiliation{Physik-Institut der Universit\"{a}t Z\"{u}rich, Winterthurerstrasse 190,
CH-8057 Z\"{u}rich, Switzerland}
\author{S.~Weyeneth}
\affiliation{Physik-Institut der Universit\"{a}t Z\"{u}rich, Winterthurerstrasse 190,
CH-8057 Z\"{u}rich, Switzerland}

\begin{abstract}
We explore the implications of Berezinskii-Kosterlitz-Thouless (BKT)
critical behavior on the two dimensional (2D) quantum
superconductor-insulator (QSI) transition driven by the tuning parameter $x$%
. Concentrating on the sheet resistance $R\left( x,T\right) $ BKT behavior
implies: an explicit quantum scaling function for $R\left( x,T\right) $
along the superconducting branch ending at the nonuniversal critical value $%
R_{c}=R\left( x_{c}\right) $; a BKT-transition line $T_{c}\left( x\right)
\propto \left( x-x_{c}\right) ^{z\overline{\nu }}$ where $z$ is the dynamic
and $\overline{\nu }$ the exponent of the zero temperature correlation
length; independent estimates of $z\overline{\nu }$, $z$ and $\overline{\nu }
$ from the $x$ dependence of the nonuniversal parameters entering the BKT
expression for the sheet resistance. To illustrate the potential and the
implications of this scenario we analyze data of Bollinger \textit{et al}.
taken on gate voltage tuned epitaxial films of La$_{2-x}$Sr$_{x}$CuO$_{4}$
that are one unit cell thick. The resulting estimates $z\simeq 2.35$ and $%
\overline{\nu }\simeq 0.63$ point to a 2D-QSI critical point where
hyperscaling, the proportionality between $d/\lambda ^{2}\left( 0\right) $
and $T_{c},$ and the correspondence between quantum phase transitions in D
and classical ones in (D+z) dimensions are violated and disorder is relevant.
\end{abstract}

\maketitle





%
%


\section{Introduction}

A variety of different materials undergo a quantum superconductor- insulator
(QSI) transition in the limit of two dimensions (2D) and zero temperature by
variation of a tuning parameter including film thickness, disorder, applied
magnetic field, and gate voltage.\cite{gold,markovic,gant,gold2} A
widespread observable to study this behavior is the temperature dependence
of the sheet resistance $R\left( x,T\right) $ taken at various values of the
tuning parameter $x$. The curves of $R(T$) at different $x$ resemble the
flow to a nearly temperature independent separatrix between superconducting
and insulating phase with sheet resistance $R_{c}=R\left( x=x_{c},T\simeq
0\right) $. This behavior implies a crossing point of the isotherms $R\left(
x\right) $ at different temperatures at $x_{c}$ which is a characteristic
feature of a quantum phase transition and in the present case of a QSI
transition. Traditionally the interpretation of experimental data taken
close to the 2D-QSI transition is based on the quantum scaling relation $%
R\left( x,T\right) =R_{c}\left( x\right) G\left( y\right) $ with $y=c\left(
x-x_{c}\right) /T^{1/z\overline{\nu }}$.\cite{sondhi,fisher1,fisher,kim} $z$
is the dynamic, $\overline{\nu }$ the critical exponent of the zero
temperature correlation length and $c$ a nonuniversal coefficient of
proportionality.\cite{gold,markovic,gant} Given sheet resistance data fits
to this scaling form yield estimates for the critical value of the tuning
parameter $x_{c}$ and the exponent product $z\overline{\nu }$, properties
which are insufficient to distinguish different models, to fix the
universality class to which the QSI transition belongs, or to clarify the
relevance of disorder .

The nature of the 2D-QSI transition has been intensely debated.\cite%
{gold,markovic,gant,gold2} The scenarios can be grouped into two classes,
fermionic and bosonic. In the fermionic case the reduction of $T_{c}$ and
the magnitude of the order parameter is attributed to a combination of
reduced density of states, enhanced Coulomb interaction and depairing due to
an increase of the inelastic electron-electron scattering rate.\cite%
{maek,finkel} The bosonic approach assumes that the fermionic degrees of
freedom can be integrated out, the mean square of the order parameter does
not vanish at $T_{c}$, phase fluctuations dominate and the reduction of $%
T_{c}$ is attributable to quantum fluctuations and in disordered systems to
randomness in addition.\cite{fisher1,fisher,weich} This scenario is closely
related to the suppression of ferroelectricity,\cite{morf} \emph{e.g}. in
SrTiO$_{3}$.\cite{alex}

Here we adopt the bosonic scenario and concentrate on systems which undergo
at finite temperature a normal state to superconductor transition with
Berezinskii-Kosterlitz-Thouless (BKT) critical behavior,\cite{bere,thoul}
originally derived for the 2D xy-model with a two component order parameter
and short range interactions. The occurrence of BKT criticality in 2D
superconductors also implies that the mean square of the order parameter
does not vanish at $T_{c}$ and with that there are, in analogy to $^{4}$He,
condensed pairs (bosons) below and uncondensed ones above $T_{c}$. In $^{4}$%
He and superconductors the order parameter is a complex scalar corresponding
to the components in the xy-model. Supposing that in superconductors the
interaction of Cooper pairs is short ranged and their effective charge is
sufficiently small the critical properties at finite temperature are then
those of the 3D-xy (bulk) and 2D-xy (thin films) models,\cite{tsbook,parks}
reminiscent to the lamda transition in bulk $^{4}$He\cite{gasp} and the
BKT-transition in thin $^{4}$He films.\cite{fino,cho,crow,oda} In this
context it is important to recognize that the existence of the
BKT-transition (vortex-antivortex dissociation instability) in $^{4}$He
films is intimately connected with the fact that the interaction energy
between vortex pairs depends logarithmic on the separation between them. As
shown by Pearl,\cite{pearl} vortex pairs in thin superconducting films
(charged superfluid) have a logarithmic interaction energy out to the
characteristic length $\lambda _{2D}$ = $\lambda ^{2}/d$, beyond which the
interaction energy falls off as $1/R$. Here $\lambda $ is the magnetic
penetration depth and $d$ the film thickness. As $\lambda _{2D}$ increases
by approaching $T_{c}$ the diamagnetism of the superconductor becomes less
important and the vortices in a clean and thin superconducting film become
progressively like those in $^{4}$He films.\cite{min,halp}

The occurrence of a 2D-QSI transition implies a line $T_{c}\left( x\right)$
of BKT transition temperatures ending at the quantum critical point at $%
x=x_{c}$ where $T_{c}\left( x=x_{c}\right) $ $=0$. It separates the
superconducting from the insulating ground state. The BKT transition is
rather special because the correlation length diverges above $T_{c}$ as $\xi
\left( x,T\right) =\xi _{0}\left( x\right) \exp \left( \left( b_{R}\left(
x\right) /2T_{c}^{1/2}\left( x\right) \right) \left( T/T_{c}\left( x\right)
-1\right) ^{-1/2}\right) $ and the sheet resistance tends to zero according
to $R\left( x,T\right) =R_{0}\left( x\right) \exp \left( -\left( b_{R}\left(
x\right) /T_{c}^{1/2}\left( x\right) \right) \left( T/T_{c}\left( x\right)
-1\right) ^{-1/2}\right) $.\cite{min,halp,tsintf} Approaching the 2D-QSI
transition quantum phase fluctuations renormalize $R_{0}\left( x\right) $, $%
b_{R}\left( x\right)$, and $T_{c}\left( x\right)$. Indeed the BKT-transition
line approaches the 2D-QSI transition as $T_{c}=\left( c\left(
x-x_{c}\right) /y_{c}\right) ^{z\overline{\nu }}$ because the quantum
scaling form $G\left( y\right) $ exhibits at the universal value $y_{c}$ of
the scaling argument a finite temperature singularity.\cite%
{fisher1,kim,tsbook} Noting that the amplitude of the BKT correlation length
$\xi _{0}\left( x\right) $ should match the divergence of the quantum
counterpart, $\xi \left( T=0\right) \propto \left( x-x_{c}\right) ^{-%
\overline{\nu }}\propto T_{c}^{-1/z}$ the exponents $z$ and $\overline{\nu }$
should emerge from the amplitude $R_{0}\left( x\right)$ in terms of $%
R_{0}\left(x\right) -R_{c}\propto \xi _{0}^{-2}\left( x\right) \propto \xi
^{-2}\left(T=0\right) \propto \left( x-x_{c}\right) ^{2\overline{\nu }%
}\propto T_{c}^{2/z}$.\cite{tsintf} Accordingly, the nonuniversal functions $%
T_{c}\left( x\right)$ and $R_{0}\left( x\right)$ entering the BKT expression
for the sheet resistance exhibit close to the QSI transition quantum
critical properties disclosing the quantum critical exponents $z\overline{%
\nu }$, $z$ and $\overline{\nu }$. The exponents $z$ and $\overline{\nu }$
are characteristic properties of the universality class to which the QSI
transition belongs. In addition, given their values the relevance of
disorder and the equivalence between quantum phase transitions in systems
with $D$ spatial dimensions and the ones of classical phase transitions in $%
\left( D+z\right)$ dimensions can be checked.\cite{sondhi,fisher1,tsbook}
The occurrence of a BKT transition line also implies: a nonuniversal
critical sheet resistance $R_{c}=R_{0}\left( x_{c}\right) $ because $%
R_{0}\left( x\right)$ is nonuniversal; an explicit form of the
superconductor branch of the quantum scaling function $G\left( c\left(
x-x_{c}\right) /T^{1/z\overline{\nu }}\right)$.

Even though BKT critical behavior is not affected by short-range correlated
and uncorrelated disorder\cite{harris,chay} the observation of this behavior
requires sufficiently homogeneous films and a tuning parameter which does
not affect the disorder. Noting that sample inhomogeneity and vortex pinning
are relevant in thickness and perpendicular magnetic field tuned
transitions, electrostatic tuning of the 2D-QSI transition using the
electric field effect appears to be more promising.\cite%
{tsintf,paren,bol,leng,bert} Indeed electrostatic tuning is not expected to
alter physical or chemical disorder, but changes the mobile carrier density.

In Sec. II we sketch the theoretical background. To illustrate the potential
and the implications of finite temperature BKT criticality on the 2D-QSI
transition we analyze in Sec. II the sheet resistance data of Bollinger
\textit{et al}.\cite{bol} taken on gate voltage tuned epitaxial films of La$%
_{2-x}$Sr$_{x}$CuO$_{4}$ that are one unit cell thick. Commenting on the
difficulties in observing the BKT features in the magnetic penetration depth
we consider the data of Bert \textit{et al}.\cite{bert} taken on the
superconducting LaAlO$_{3}$/SrTiO$_{3}$ interface.

\section{Theoretical background}

Continuous quantum-phase transitions (QPT) are transitions at zero
temperature in which the ground state of a system is changed by varying a
parameter of the Hamiltonian.\cite{sondhi,tsbook,parks} The quantum
superconductor- insulator transitions (QSI) in two-dimensional (2D) systems
tuned by disorder, film thickness, magnetic field or with the electrostatic
field effect are believed to be such transitions.\cite%
{gold,markovic,gant,tsintf,tsbook,bol,parks} Traditionally the
interpretation of experimental data taken close to the 2D-QSI transition is
based on the quantum scaling relation,\cite{sondhi,fisher1,fisher,kim}%
\begin{equation}
\frac{R\left( x,T\right) }{R_{c}}=G\left( y\right) ,y=\frac{c\left\vert
x-x_{c}\right\vert }{T^{1/z\overline{\nu }}}.  \label{eq1}
\end{equation}%
where $R$ is the resistance per square, $R_{c}$ the limiting ($x\rightarrow
x_{c}$ and $T\rightarrow 0$) resistance. $x$ denotes the tuning parameter
and $c$ a nonuniversal coefficient of proportionality. $G\left( y\right) $
is a universal scaling function of its argument such that $G\left(
y=0\right) =1$. In addition, $z$ is the dynamic and $\overline{\nu }$ the
critical exponent of the correlation length, supposed to diverge as
\begin{equation}
\xi \left( T=0\right) =\overline{\xi _{0}}\left\vert x-x_{c}\right\vert ^{-%
\overline{\nu }}.  \label{eq1a}
\end{equation}%
The critical sheet resistance $R_{c}$ separating the superconducting and
insulating ground states is determined from the isothermal sheet resistance
at the crossing point in $R\left( T\right) $ \textit{vs}. tuning parameter $%
x $ at $x_{c}$. The existence of such a crossing point, remaining
temperature independent in the zero temperature limit, is the signature of a
QPT. The data for $R\left( x,T\right) $ plotted \textit{vs}. $\left\vert
x-x_{c}\right\vert /T^{1/\overline{\nu }z}$ should then collapse onto two
branches joining at $R_{c}$. The lower branch stems from the superconducting
$\left( x-x_{c}>0\right) $ and the upper one from the insulating phase $%
\left( x-x_{c}<0\right) $. This scaling form follows by noting that the
divergence of the zero temperature correlation length, $\xi \left(
T=0\right) =\xi _{0}\left\vert \delta \right\vert ^{-\overline{\nu }}$, is
at finite temperature limited by the length $L_{T}\propto T^{-1/z}$.\cite%
{sondhi} Thus $G\left( y\right) $ is a finite-size scaling function because $%
y\propto \left[ L_{T}/\xi \left( T=0\right) \right] ^{1/\overline{\nu }%
}\propto $ $\left\vert x-x_{c}\right\vert /T^{1/\overline{\nu }z}$.
Supposing that there is a line of finite temperature phase transitions $%
T_{c}\left( x\right) $ ending at the quantum critical point $T_{c}\left(
x=x_{c}\right) =0$ the quantum scaling form (\ref{eq1}) exhibits at the
universal value $y_{c}$ of the scaling argument a finite temperature
singularity.\cite{fisher1,kim,tsbook} The phase transition line is then
fixed by%
\begin{equation}
T_{c}=\left( \frac{c\left\vert x-x_{c}\right\vert }{y_{c}}\right) ^{z%
\overline{\nu }}\text{.}  \label{eq2}
\end{equation}%
Otherwise one expects that sufficiently homogeneous 2D superconductors
exhibit at the superconductor to normal state transition BKT critical
behavior.\cite{tsintf,bere,thoul} Note that there is the Harris criterion,%
\cite{harris} stating that short-range correlated and uncorrelated disorder
is irrelevant at the unperturbed critical point, provided that $\nu >2/D$,
where $D$ is the dimensionality of the system and $\nu $ the critical
exponent of the finite-temperature correlation length. With $D=2$ and $\nu
=\infty $ , appropriate for the BKT transition, \cite{bere,thoul} this
disorder should be irrelevant. \ Given a BKT superconductor to normal state
transition the sheet resistance scales for $T\gtrsim T_{c}\left( x\right)
\geq 0$ as\cite{min,halp,tsintf}
\begin{equation}
\frac{R\left( x,T\right) }{R_{0}\left( x\right) }=\exp \left( -\frac{%
b_{R}\left( x\right) }{T_{c}^{1/2}\left( x\right) \left( T/T_{c}\left(
x\right) -1\right) ^{1/2}}\right) ,  \label{eq3}
\end{equation}%
allowing to probe the characteristic BKT correlation length\cite%
{bere,thoul,halp,tsintf}%
\begin{equation}
\frac{\xi \left( x,T\right) }{\xi _{0}\left( x\right) }=\exp \left( \frac{%
b_{R}\left( x\right) }{2T_{c}^{1/2}\left( x\right) \left( T/T_{c}\left(
x\right) -1\right) ^{1/2}}\right) .  \label{eq4}
\end{equation}%
While $R_{0}\left( x\right) $, $b_{R}\left( x\right) $, and $T_{c}\left(
x\right) $ depend on the tuning parameter and are subject to quantum
fluctuations the characteristic BKT form of the correlation length and with
that of the sheet resistance applies for any $T\gtrsim T_{c}\left( x\right)
\geq 0$. Through standard arguments (see, e.g., Ref. \cite{hertz}) quantum
mechanics does not modify universal finite temperature properties. $%
b_{R}\left( x\right) $ is given by\cite{thoul,tsintf}
\begin{equation}
b_{R}\left( x\right) =\widetilde{b}_{R}T_{c}^{-1/2},  \label{eq5}
\end{equation}%
with $\widetilde{b}_{R}=4\pi /b$. The parameter $b$ is expected to remain
constant in the low $T_{c}$ regime.\cite{tsintf} It is related to the vortex
core energy $E_{c}$ in terms of $b=f\left( E_{c}/k_{B}T_{c}\right) $\cite%
{steel} and controls below the Nelson-Kosterlitz jump the temperature
dependence of the magnetic penetration depth in terms of $\left( \lambda
\left( T_{c}\right) /\lambda \left( T\right) \right) ^{2}=1+\left(
b/4\right) \left( T/T_{c}-1\right) ^{1/2}$.\cite{nelson} The amplitude of
the BKT correlation length $\xi _{0}\left( x\right) $ is proportional to the
vortex core radius\cite{fino} known to increase with reduced $T_{c}$.\cite%
{cho,oda} Indeed, the zero temperature correlation length $\xi \left(
T=0\right) $ diverges as $\xi \left( T=0\right) \propto \left(
x-x_{c}\right) ^{-\overline{\nu }}$ (Eq. (\ref{eq1a}) and combined with $%
T_{c}$ $\propto \left( x-xc\right) ^{z\overline{\nu }}$(Eq. (\ref{eq2})) we
obtain $\xi \left( T=0\right) \propto \left( x-x_{c}\right) ^{-\overline{\nu
}}\propto T_{c}^{-1/z}$. Noting that $R_{0}\left( x\right) $ approaches $%
R_{c}$ from above the scaling relation%
\begin{equation}
R_{0}\left( x\right) -R_{c}\propto \xi ^{-2}\left( T=0\right) \propto \left(
x-x_{c}\right) ^{2\overline{\nu }}\propto T_{c}^{2/z},  \label{eq6}
\end{equation}%
should apply,\cite{tsintf} making the determination of the exponents $%
\overline{\nu }$ and $z$ possible. Consistency requires that the resulting $z%
\overline{\nu }$ agrees with the estimate obtained from the critical BKT
line $T_{c}\left( x\right) \propto \left( x-x_{c}\right) ^{z\overline{\nu }}$%
. Other implications concern the universality class of the 2D-QSI transition
and the relevance of disorder. Given estimates for $\overline{\nu }$ and $z$
\ the equivalence between quantum phase transitions in clean systems with $D$
spatial dimensions and the ones of classical phase transitions in $\left(
D+z\right) $ dimensions can be checked. The fate of a clean critical point
under the influence of disorder is controlled by the Harris criterion:\cite%
{harris,chay} If the zero temperature correlation length critical exponent
fulfills the inequality $\overline{\nu }\geq 2/D$ the disorder does not
affect the critical behavior. If the Harris criterion is violated, $%
\overline{\nu }<2/D$, the generic result is a new critical point with
conventional power law scaling but new exponents which fulfill the Harris
criterion. Another option is that the disorder destroys the QSI transition.

Finally, given a BKT transition line expression (\ref{eq3}) the sheet
resistance given by Eq. (\ref{eq4}) transforms with Eq. (\ref{eq5}) and the
scaling variable $y=c\left\vert x-x_{c}\right\vert /T^{1/z\overline{\nu }}$
to%
\begin{equation}
\ln \left( \frac{R\left( x,T\right) }{R_{0}\left( x\right) }\right) =-%
\widetilde{b}_{R}\left( \left( \frac{y_{c}}{y}\right) ^{z\overline{\nu }%
}-1\right) ^{-1/2},  \label{eq7}
\end{equation}%
valid foe $y\leq y_{c}$. Close to quantum criticality where $R_{0}\left(
x\right) \simeq R_{c}$ reduces to
\begin{eqnarray}
\ln \left( \frac{R\left( x,T\right) }{R_{c}}\right) &=&-\widetilde{b}%
_{R}\left( \left( \frac{y_{c}}{y}\right) ^{z\overline{\nu }}-1\right) ^{-1/2}
\nonumber \\
&=&\ln \left( G\left( y\right) \right) .  \label{eq8}
\end{eqnarray}%
These relations are explicit forms of the quantum scaling function $G\left(
y\right) $ applicable to the superconductor branch. They reveal that the
critical sheet resistance $R_{0}\left( x\rightarrow x_{c}\right) =R_{c}$ is
the endpoint of a nonuniversal function and accordingly nonuniversal.
Another implication concerns the universality class of the 2D-QSI
transition. Supposing that the equivalence between quantum phase transitions
in clean systems with $D$ spatial dimensions and the ones of classical phase
transitions in $\left( D+z\right) $ dimensions applies, the 2D-QSI
transition at the endpoint of a BKT line $T_{c}\left( x\right) $ should
belong to the finite temperature $\left( 2+z\right) -xy$ universality class.
$xy$ denotes an order parameter with two components, including the complex
scalar, $\Psi =\left\vert \Psi \right\vert \exp \left( i\varphi \right) $,
of a superconductor.\cite{sondhi,tsbook}

The BKT - theory of thermally-excited vortex-antivortex pairs also predicts
a super-to-normal state phase transition marked by the Nelson-Kosterlitz
jump, a discontinuous drop in superfluid density from\cite{nelson},%
\begin{equation}
\frac{d}{\lambda ^{2}\left( T_{c}^{-}\right) }=\frac{32\pi ^{2}k_{B}T_{c}}{%
\Phi _{0}^{2}}\simeq 1.02T_{c},  \label{eq9}
\end{equation}%
to zero. The numerical relationship applies for $d/\lambda ^{2}\left(
T_{c}^{-}\right) $ in cm$^{-1}$ and $T_{c}$ in K. $d$ denotes the thickness
of the 2D system, $\lambda $ the in-plane magnetic penetration depth, and $%
\Phi _{0}=hc/2e$. In addition there is the prediction that $d/\lambda
^{2}\left( T=0\right) $, a measure of the phase stiffness, scales near the
endpoint of the BKT transition line as\cite{tsbook,parks,fisher1,kim}
\begin{equation}
\frac{d}{\lambda ^{2}\left( T=0\right) }=\frac{16\pi ^{3}k_{B}T_{c}}{\Phi
_{0}^{2}}Q_{2}\simeq 1.6Q_{2}T_{c},  \label{eq10}
\end{equation}%
provided that $D+z$ is below the upper critical dimension $D_{u}$ where
hyperscaling holds. Since $D_{u}=4$ in the $D$-xy-model the validity of Eq. (%
\ref{eq10}) is in $D=2$ restricted to $z<2$. Relation (\ref{eq10}) is the
quantum counterpart \ of the Nelson-Kosterlitz relation (\ref{eq7}). $Q_{2}$
is a dimensionless critical amplitude with the lower bound\cite{tsbook,parks}
\begin{equation}
Q_{2}\geq 2/\pi ,  \label{eq11}
\end{equation}%
dictated by the characteristic temperature dependence of $d/\lambda
^{2}\left( T\right) $ below the Nelson-Kosterlitz jump (Eq. (\ref{eq9})).%
\cite{tsbook,parks} Combining Eqs. (\ref{eq9}) and (\ref{eq10}) we obtain
\begin{equation}
\left( \frac{\lambda \left( T=0\right) }{\lambda \left( T_{c}\right) }%
\right) ^{2}=\frac{\rho _{s}\left( T=0\right) }{\rho _{s}\left( T_{c}\right)
}=\frac{\pi }{2}Q_{2},  \label{eq11a}
\end{equation}%
where $\rho _{s}$ is the superfluid density. The superfluid transition
temperature $T_{c}$ as a function of the superfluid density $\rho _{s}\left(
T=0\right) $ has been measured in $^{4}$He films for transition temperatures
ranging from $0.3$ to $1$ K by Crowell \textit{et al}.\cite{crow} They
studied $^{4}$He films adsorbed in two porous glasses, aerogel and Vycor,
using high-precision torsional oscillator and dc calorimetry techniques. The
investigation focused on the onset of superfluidity at low temperatures as
the $^{4}$He coverage is increased. Their data yields $\rho _{s}\left(
T=0\right) \simeq 15.3T_{c}$ with $\rho _{s}$ in $\mu $moles/m$^{2}$ and $%
T_{c}$ in K. Combined with the BKT transition line $\rho _{s}\left(
T_{c}=0\right) =8.73T_{c}$ we obtain $Q_{2}\simeq 1.12$.

\section{COMPARISON WITH EXPERIMENT}

\begin{figure}[tbh]
\includegraphics[width=1.0\linewidth]{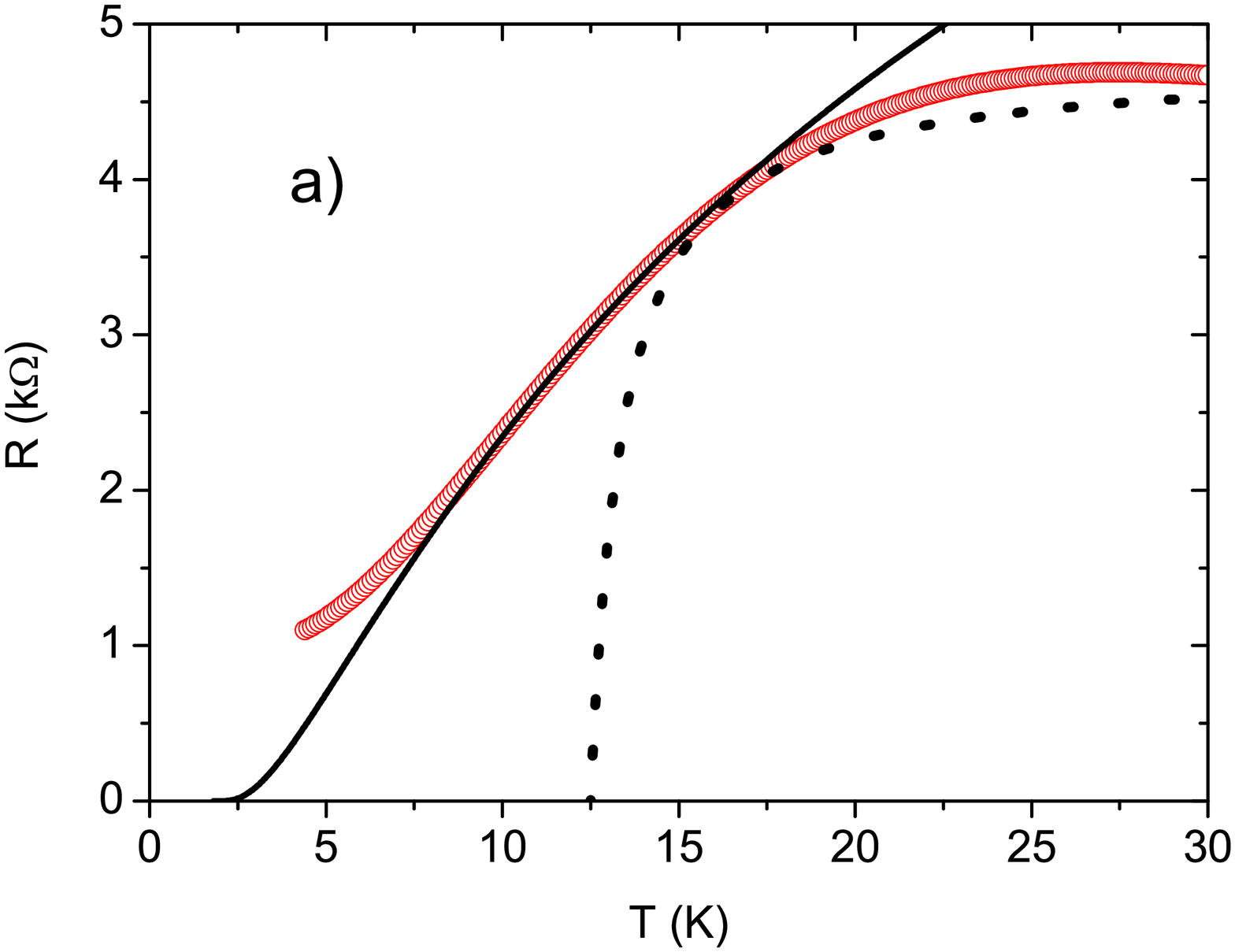} \vspace{-0.5cm} %
\includegraphics[width=1.0\linewidth]{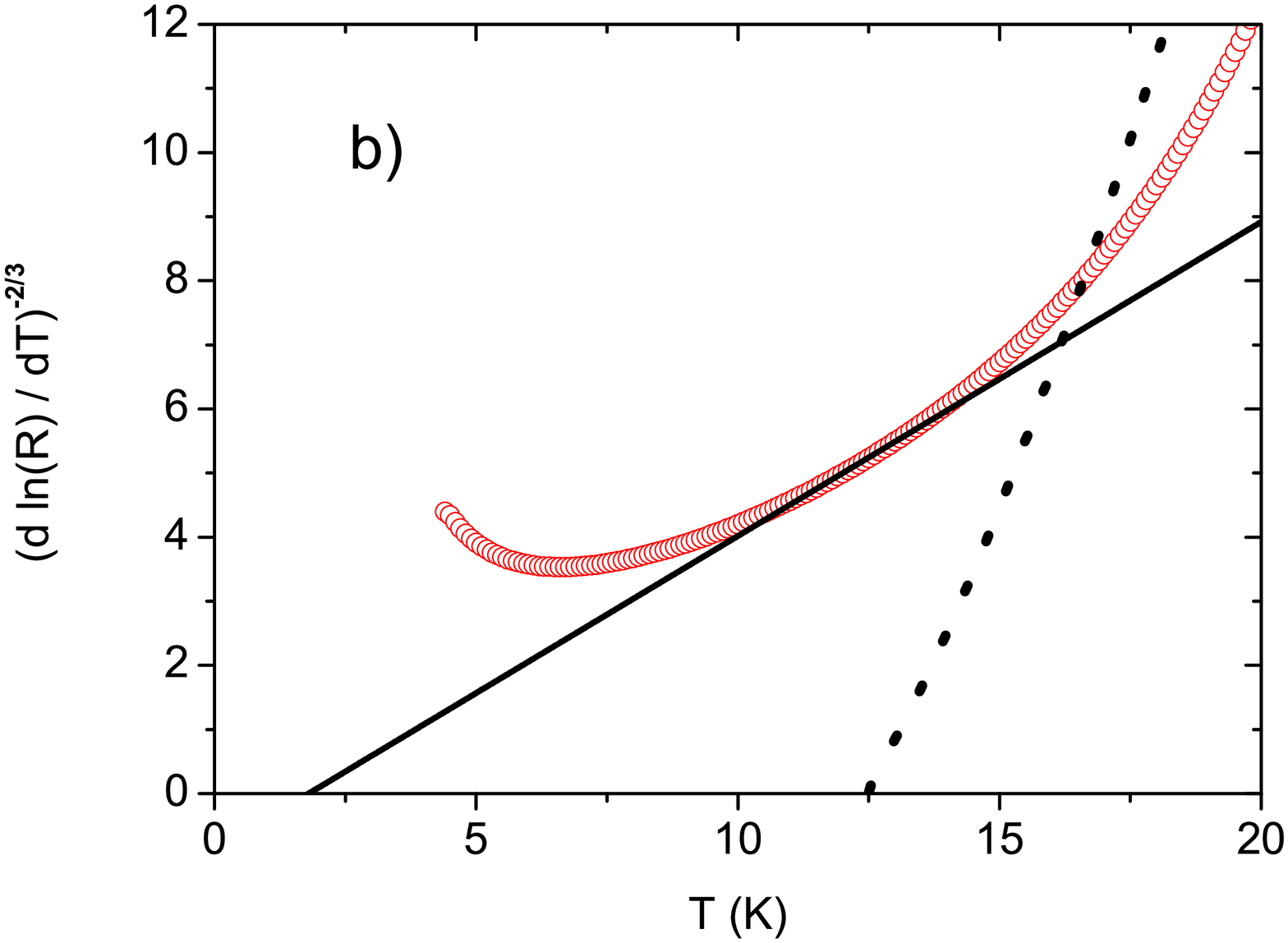} \vspace{-0.5cm}
\caption{(color online) a) Sheet resistance $R\left( T\right) $ for $%
V_{g}=-1.5$ V taken from Bollinger \textit{et al}.\protect\cite{bol} The
dotted curve is a fit to the Aslamosov-Larkin expression yielding the $%
T_{c0}\simeq 12.5$ K. The solid line is the BKT behavior Eq. (\protect\ref%
{eq3}) with $R_{0}=18$ k$\Omega $ and the $T_{c}$, $b_{R}$ values derived in
Fig. \protect\ref{fig1}b. b) $\left( d\ln \left( R\right) /dT\right) ^{-2/3}$
\textit{vs}. $T$ derived from the data shown in Fig. \protect\ref{fig1}a.
The dotted line indicates the AL-behavior and the solid one Eq. (\protect\ref%
{eq11a}) with $T_{c}=1.8$ K and $\left( 2/b_{R}\left( T_{c}\right) \right)
^{2/3}$ $=0.49$ (K$^{-1/3}$)}
\label{fig1}
\end{figure}

To illustrate the potential and the implications of the outlined BKT
scenario we analyze next the data of Bollinger \textit{et al}.\cite{bol}
taken on epitaxial films of La$_{2-x}$Sr$_{x}$CuO$_{4}$ that are one unit
cell thick. Very large electric fields and the associated changes in surface
carrier density enabled shifts in the midpoint transition temperature $T_{c}$
by up to $30$ K. Hundreds of resistance versus temperature and carrier
density curves were recorded and shown to collapse onto a single function,
as the quantum scaling form (Eq. (\ref{eq1})) for a 2D-QSI transition
predicts. The observed critical resistance is close to the quantum
resistance for pairs, $R_{Q}=h/4e^{2}$ $\simeq 6.45$ k$\Omega $. Our
starting point is the temperature dependence of the sheet resistance taken
at various gate voltages $V_{g}$ where a BKT transition is expected to
occur. As an example we depicted in Fig. \ref{fig1}a the data for $%
V_{g}=-1.5 $ V. \ The observation of BKT-behavior requires that the data
extend considerably below the mean-field transition temperature $T_{c0}$. We
estimated it with the aid of the Aslamosov-Larkin (AL) expression\cite%
{aslamo} for the conductivity, $\sigma \left( T,V_{g}\right) =\sigma
_{n}\left( V_{g}\right) +\sigma _{0}/\left( T/T_{c0}-1\right) $, with $%
\sigma _{0}=\pi e^{2}/8h\simeq 1.52\times 10^{-5}$ $\Omega ^{-1}$, where
Gaussian fluctuations are taken into account and $T_{c0}$ is the mean-field
transition temperature. The resulting temperature dependence is included in
Fig. \ref{fig1}a. It clearly reveals that the data extend considerably below
$T_{c0}\simeq 12.5$ K. To establish and characterize BKT behavior below $%
T_{c0}$ we invoke Eq. (\ref{eq3}) and
\begin{equation}
\left( \frac{d\ln \left( R\right) }{dT}\right) ^{-2/3}=\left( \frac{2}{%
b_{R}\left( T_{c}\right) }\right) ^{2/3}\left( T-T_{c}\left( x\right)
\right) .  \label{eq11b}
\end{equation}%
As indicated in Fig. \ref{fig1}b this relation is used to fix $b_{R}\left(
T_{c}\right) $ and $T_{c}\left( x\right) $ while $R_{0}\left( x\right) $ is
estimated by adjusting Eq. (\ref{eq3}) with given $b_{R}\left( T_{c}\right) $
and $T_{c}\left( x\right) $ to the sheet resistance data. Comparing the data
with the respective lines we observe that the BKT regime is attained and
that hat the BKT $T_{c}$ is almost an order of magnitude lower than the
mean-field counterpart. This uncovers a BKT transition from uncondensed to
condensed Cooper \ pairs driven by strong phase fluctuations. On the other
hand it is important to recognize that agreement with BKT-criticality is
established in a temperature window only. Its upper bound reflects the
crossover from BKT- to AL-behavior while the lower bound stems from the
rounded BKT-transition. Precursors of this phenomenon are clearly visible in
Fig. \ref{fig1} around $7$ K. Here the correlation length is prevented to
grow beyond a limiting length $L$, \textit{i.e.} the linear extent of the
homogeneous domains. As a result, a finite size effect and with that a
rounded transition occurs.\cite{tsintf} Because the BKT correlation length
does not exhibit the usual power law divergence of the correlation length as
$T_{c}$ is approached, it is particularly susceptible to such finite-size
effect.

\begin{figure}[tbh]
\includegraphics[width=1.0\linewidth]{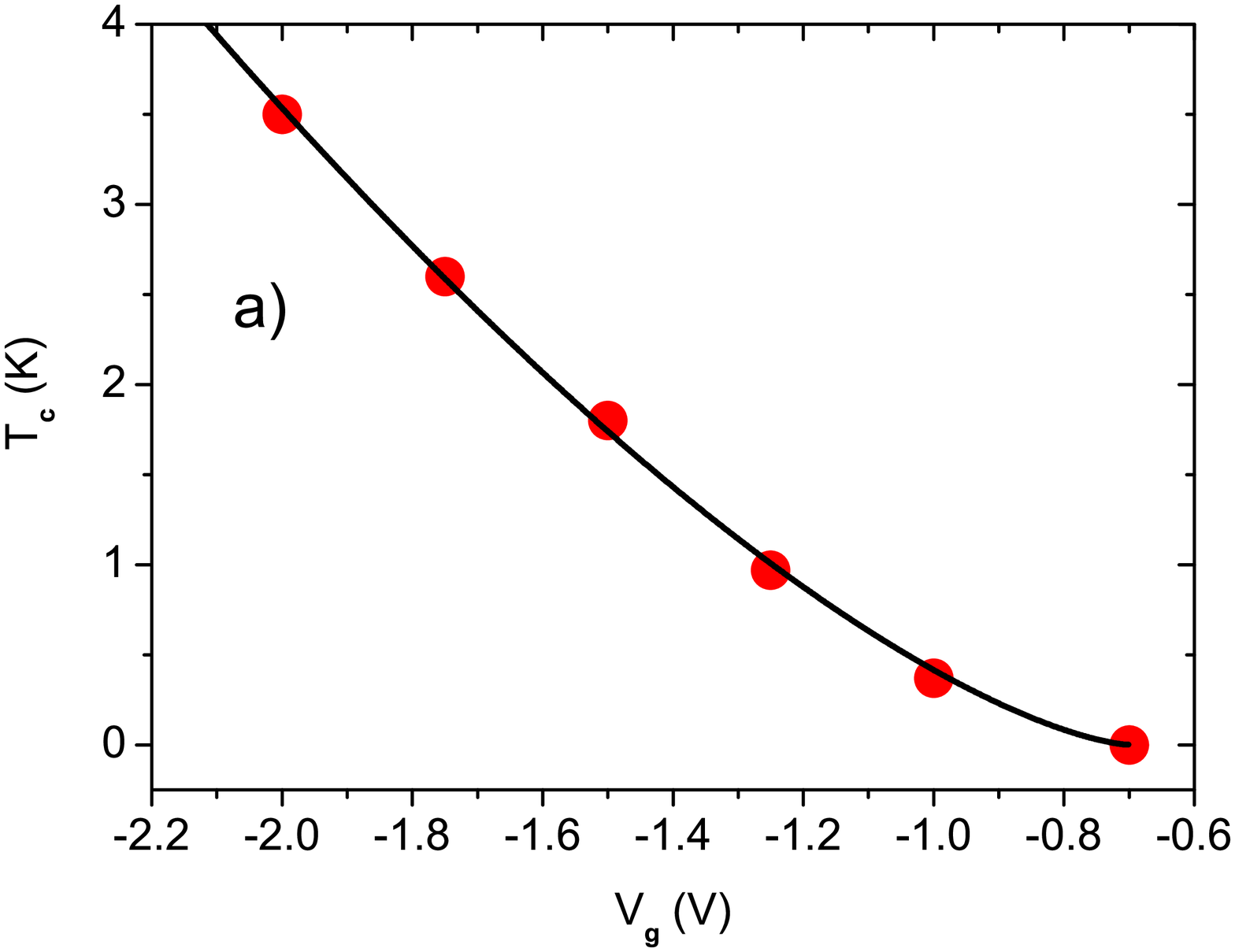} \vspace{-0.5cm} %
\includegraphics[width=1.0\linewidth]{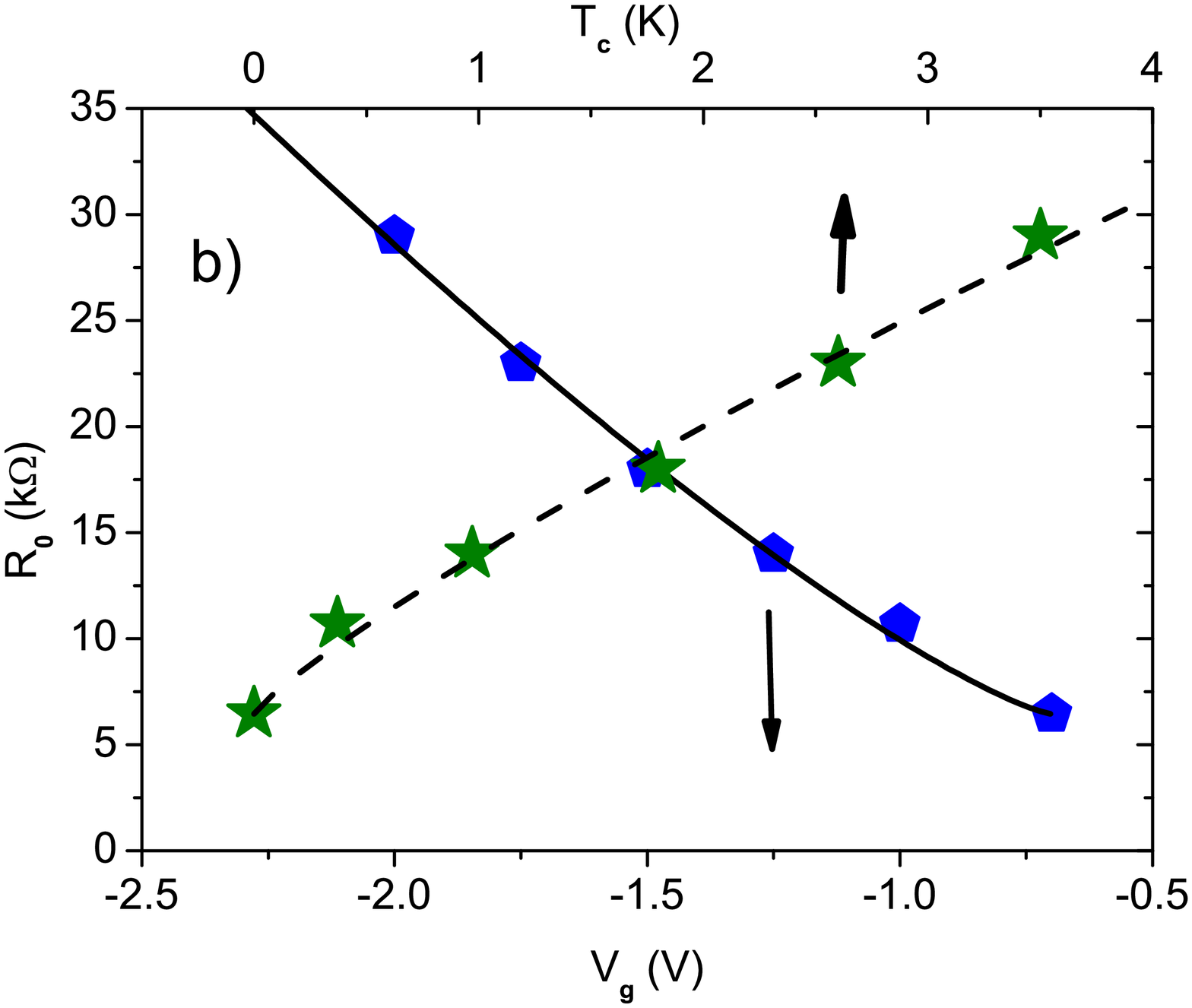} \vspace{-0.5cm}
\caption{(color online) a) Phase transition line $T_{c}\left( V_{g}\right) $
derived from the sheet resistance data of Bollinger \textit{et al}.%
\protect\cite{bol} using Eq. (\protect\ref{eq11b}) . The solid line is a
least square fit yielding $T_{c}=2.41\left( V_{gc}-V_{g}\right) ^{z\overline{%
\protect\nu }}$ (K) and $z\overline{\protect\nu }=1.46$ with $V_{gc}=-0.7$
V. b) $R_{0}\left( V_{g}\right) $ derived from the sheet resistance data of
Bollinger \textit{et al}.\protect\cite{bol} using Eq. (\protect\ref{eq3})
with given $T_{c}\left( V_{g}\right) $ and $b_{R}\left( T_{c}\right) $; $%
R_{0}\left( T_{c}\right) $ derived from $R_{0}\left( V_{g}\right) $ using $%
T_{c}\left( V_{g}\right) $. The solid line is $R_{0}\left( V_{g}\right)
=R_{c}+\left( V_{gc}-V_{g}\right) ^{2\overline{\protect\nu }}$ (k$\Omega $)
with $R_{c}=6.45$ k$\Omega $, $V_{gc}=-0.7$ V and $\overline{\protect\nu }%
=0.63$. The dashed line is $R_{0}\left( T_{c}\right) =R_{c}+7.52T_{c}^{2/z}$
(k$\Omega $) with $z=2.35$.}
\label{fig2}
\end{figure}

\begin{figure}[tbh]
\includegraphics[width=1.0\linewidth]{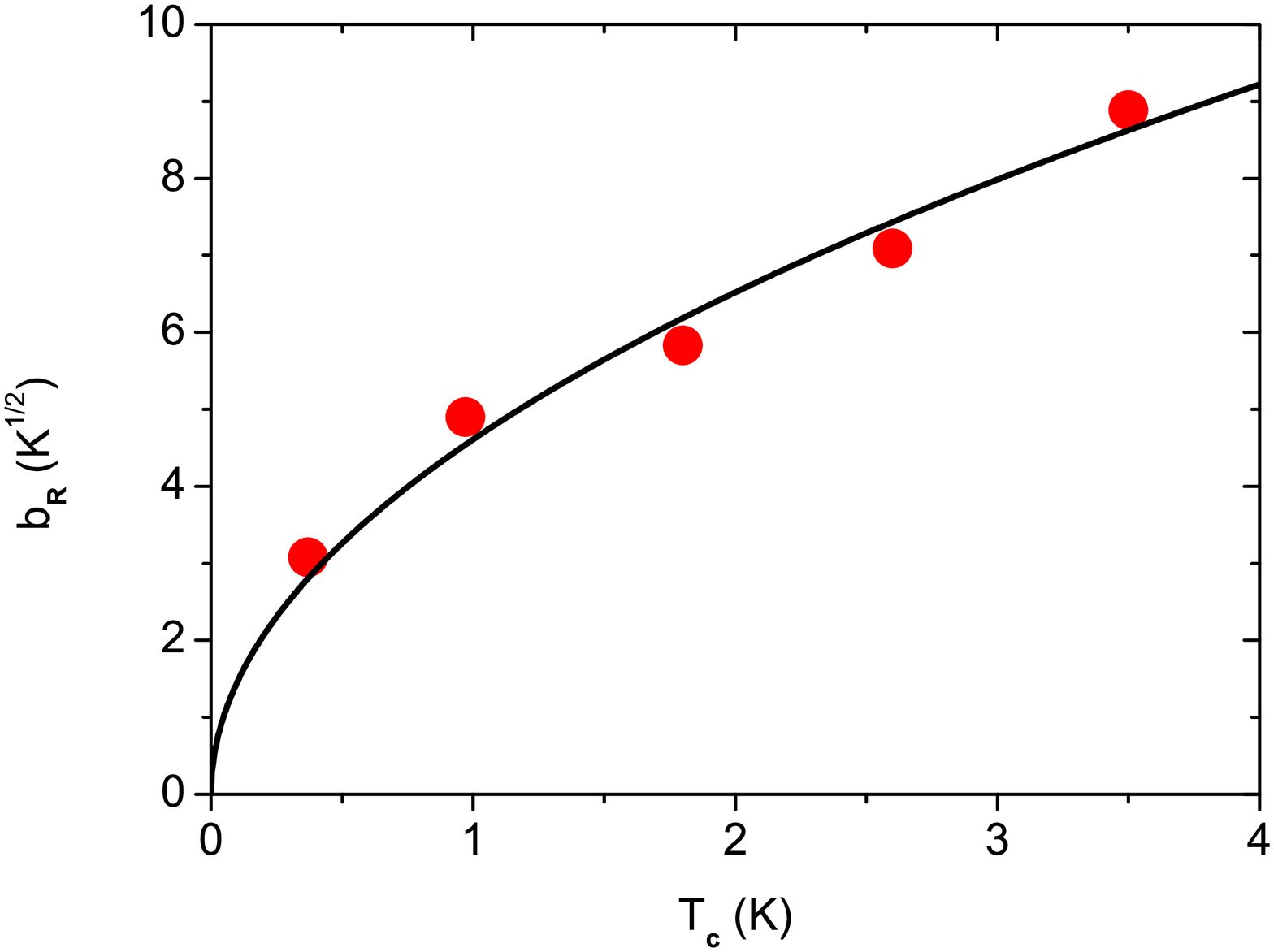} \vspace{-0.5cm}
\caption{(color online) Estimates for $b_{R}\left( T_{c}\right) $ derived
from the sheet resistance data from Bollinger \textit{et al}.\protect\cite%
{bol} using Eq. (\protect\ref{eq11a}). The solid line is $b_{R}\left(
T_{c}\right) =\widetilde{b}_{R}T_{c}^{1/2}$ (K$^{1/2}$) with $\widetilde{b}%
_{R}=4.61$ in agreement with Eq. (\protect\ref{eq5}).}
\label{fig3}
\end{figure}

\begin{figure}[tbh]
\includegraphics[width=1.0\linewidth]{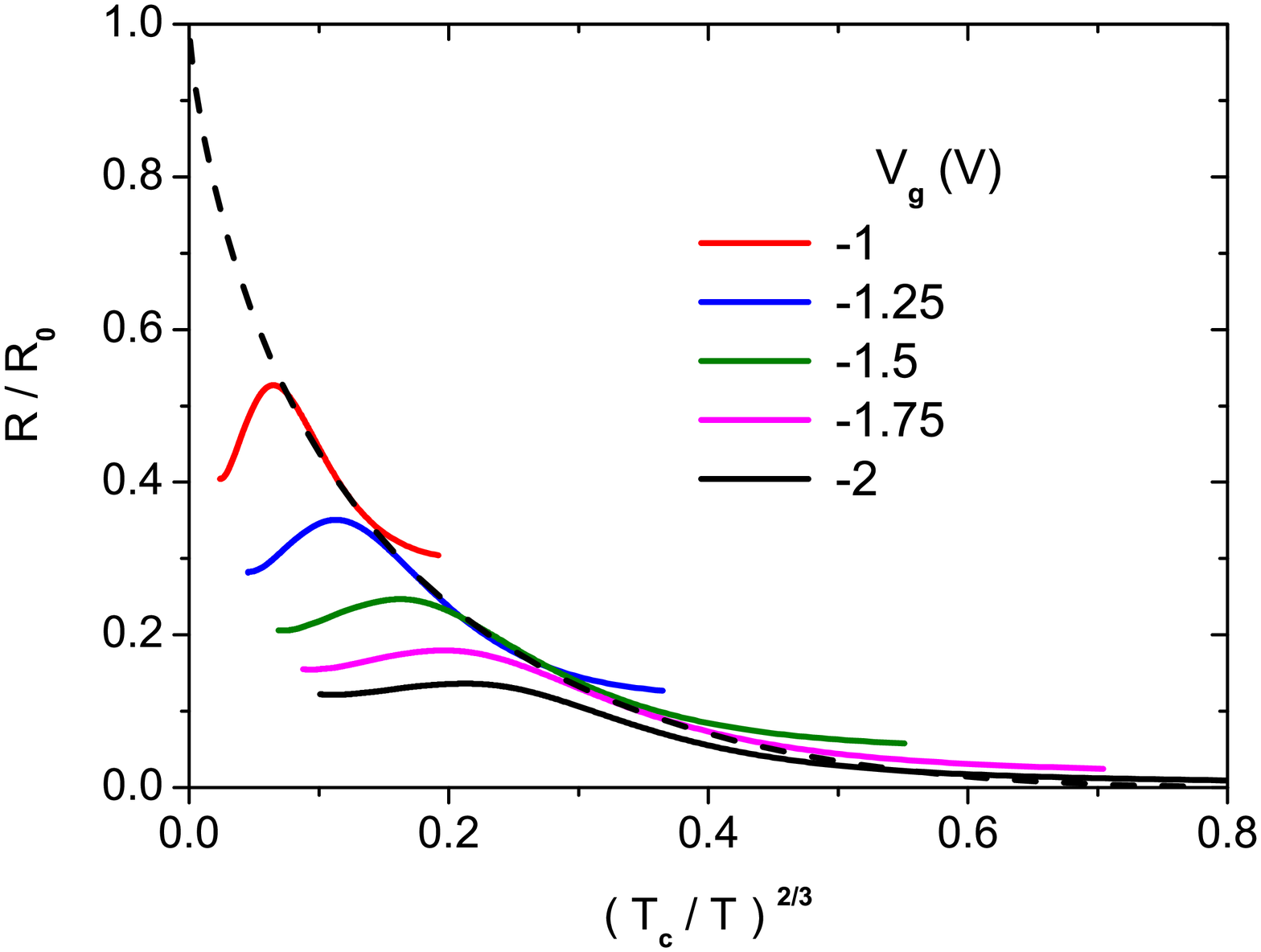} \vspace{-0.5cm}
\caption{(color online) $R\left( V_{g},T\right) /R_{0}\left( V_{g}\right) $
\textit{vs}.$\left( T_{c}\left( V_{g}\right) /T\right) ^{2/3}=y/y_{c}$
corresponding to the BKT quantum scaling function $G\left( y\right) $ (Eq. (%
\protect\ref{eq7})). $R\left( V_{g},T\right) $ is taken from Bollinger
\textit{et al}.\protect\cite{bol} in the temperature range from $4.4$ K to $%
100$ K. The respective values for $R_{0}\left( x\right) $, $T_{c}\left(
V_{g}\right) $ and $\widetilde{b}_{R}$ are shown in Figs. \protect\ref{fig2}
and \protect\ref{fig3}. The dashed line is the critical BKT behavior given
by Eq. (\protect\ref{eq3}) with $\widetilde{b}_{R}=b_{R}/T_{c}^{1/2}=4.55$.}
\label{fig4}
\end{figure}

To explore the quantum critical behavior latent in $T_{c}\left( V_{g}\right)
$, $R_{0}\left( V_{g}\right) $ and $b_{R}\left( T_{c}\right) $ we performed
this analysis of the temperature dependence of the sheet resistance for
additional gate voltages. To fix the critical gate voltage $V_{gc}$ we used
the empirical gate voltage dependence of the number of mobile holes $x$ per
one formula unit of Bollinger \textit{et al}.\cite{bol} yielding $x\left(
V_{g}\right) =x_{c}+0.012(V_{gc}-V_{g})$ down to $V_{g}=-2$V with $%
x_{c}=0.0605$ and $V_{gc}\simeq -0.7$ V. The results $T_{c}\left(
V_{g}\right) $, $R_{0}\left( V_{g}\right) $ are shown in Fig. \ref{fig2}a
and Fig. \ref{fig2}b while $b_{R}\left( T_{c}\right) $ is depicted in Fig. %
\ref{fig3}. As can be seen in Fig. \ref{fig2}a the BKT transition line
differs substantially from the so called superconducting dome behavior
observed in bulk cuprate superconductors. It is approximately given by $%
T_{c}\left( x\right) /T_{c\max }=1-82.6\left( x-x_{m}\right) ^{2}$ where $%
T_{c\max }\left( x_{m}=0.16\right) $ is the maximum $T_{c}$.\cite{pres}
Close to the QSI transition where even bulk cuprate superconductors become
essentially 2D\cite{parks} it reduces to $T_{c}\left( x\right) /T_{c\max
}\simeq 10.3\left( x-x_{c}\right) $ with $x_{c}\simeq 0.049$ and suggests $z%
\overline{\nu }\simeq 1$. In any case it differs substantially from the BKT
line shown in Fig. \ref{fig2}a yielding the estimate
\begin{equation}
z\overline{\nu }\simeq 1.46,  \label{eq12}
\end{equation}%
in agreement with the value $z\overline{\nu }\simeq 1.5$ derived by
Bollinger \textit{et al}.\cite{bol} using the quantum scaling approach. In
contrast to this, from the nonuniversal parameters $R_{0}\left( V_{g}\right)
$ and $R_{0}\left( T_{c}\right) $ shown in Fig. \ref{fig2}b we derive with
Eq. (\ref{eq6}) in addition%
\begin{equation}
\overline{\nu }\simeq 0.63,z\simeq 2.35,z\overline{\nu }=1.48.  \label{eq13}
\end{equation}%
As these exponents satisfy the inequality $\left( 2+z\right) \overline{\nu }%
\geq 2$ we use the correct scaling argument, because $\delta =\left( \mu
-\mu _{c}\right) \propto \left( x-x_{c}\right) \propto \left(
V_{gc}-V_{g}\right) $ where $\mu $ denotes the schemical potential.\cite%
{fisher1} The agreement between these $z\overline{\nu }$ values confirms the
applicability of the scaling relation (\ref{eq6}). Similarly, the
nonuniversal parameter $b_{R}\left( x\right) $ exhibits according to Fig. %
\ref{fig3} the expected $T_{c}$ dependence (Eq. (\ref{eq5})). Noting that $%
D+z\simeq 2+z\simeq 4.35$ exceeds the upper critical dimension $D_{u}=4$ the
critical exponent of the zero temperature correlation length $\overline{\nu }
$ should adopt its mean-field value $\overline{\nu }=1/2$. However the fate
of this clean critical point under the influence of disorder is controlled
by the Harris criterion.\cite{harris,chay} If the inequality \ $\overline{%
\nu }\geq 2/D$ is fulfilled, the disorder does not affect the critical
behavior. If the Harris criterion is violated (\ $\overline{\nu }<2/D$), the
generic result is a new critical point with conventional power law scaling
but new exponents which fulfill \ $\overline{\nu }<2/D$. Since $\overline{%
\nu }=1/2$ violates this inequality disorder is relevant and drives the
system from the mean-field to an other critical point with different
critical exponents as our estimate $\overline{\nu }\simeq 0.63$, fulfilling $%
\overline{\nu }<2/D=1$, uncovers. The resulting 2D-QSI transition with $%
\overline{\nu }\simeq 0.63$ and $z\simeq 2.35$ violates then the equivalence
between quantum phase transitions in systems with $D$ spatial dimensions and
the ones of classical phase transitions in $\left( D+z\right) $ dimensions.
In addition the proportionality between $d/\lambda ^{2}\left( 0\right) $ and
$T_{c}$ (Eq. (\ref{eq10})) valid below the upper critical dimension $D_{u}=4$
does no longer hold because $\left( D+z\right) \simeq 4.35$ is above $D_{u}=4
$. In fact magnetic penetration depth measurements taken on underdoped high
quality YBa$_{2}$Cu$_{3}$O$_{6+x}$ single crystals revealed $T_{c}\propto
\left( d/\lambda ^{2}\left( 0\right) \right) ^{0.61}$.\cite{liang} Another
option is that the disorder destroys the QSI transition. Given the evidence
for a rounded BKT transition (see Fig. \ref{fig1}) and the missing data
close to the QSI transition (see Fig. \ref{fig4}) further studies are
required to elucidate this option.

\begin{figure}[tbh]
\includegraphics[width=1.0\linewidth]{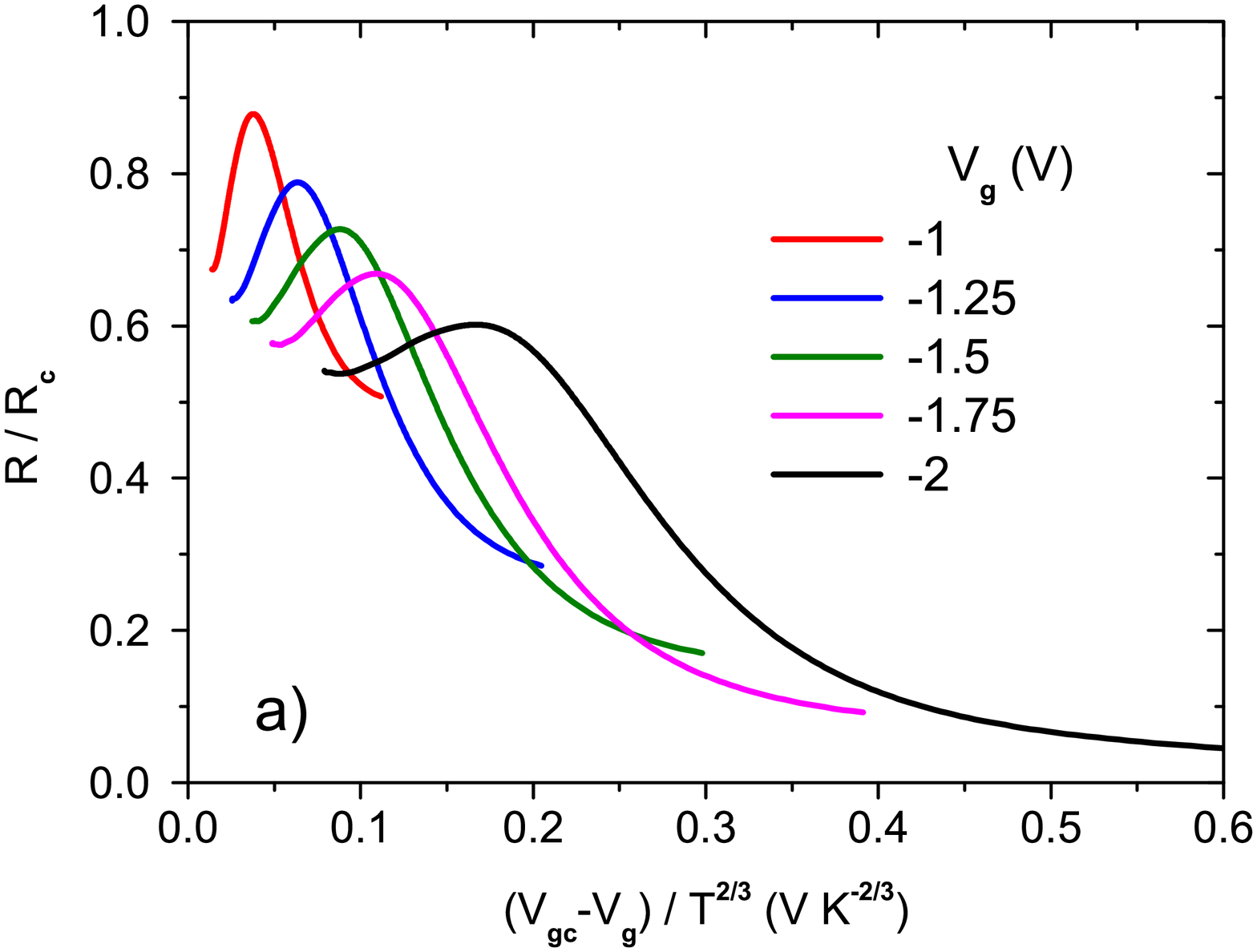} \vspace{-0.5cm} %
\includegraphics[width=1.0\linewidth]{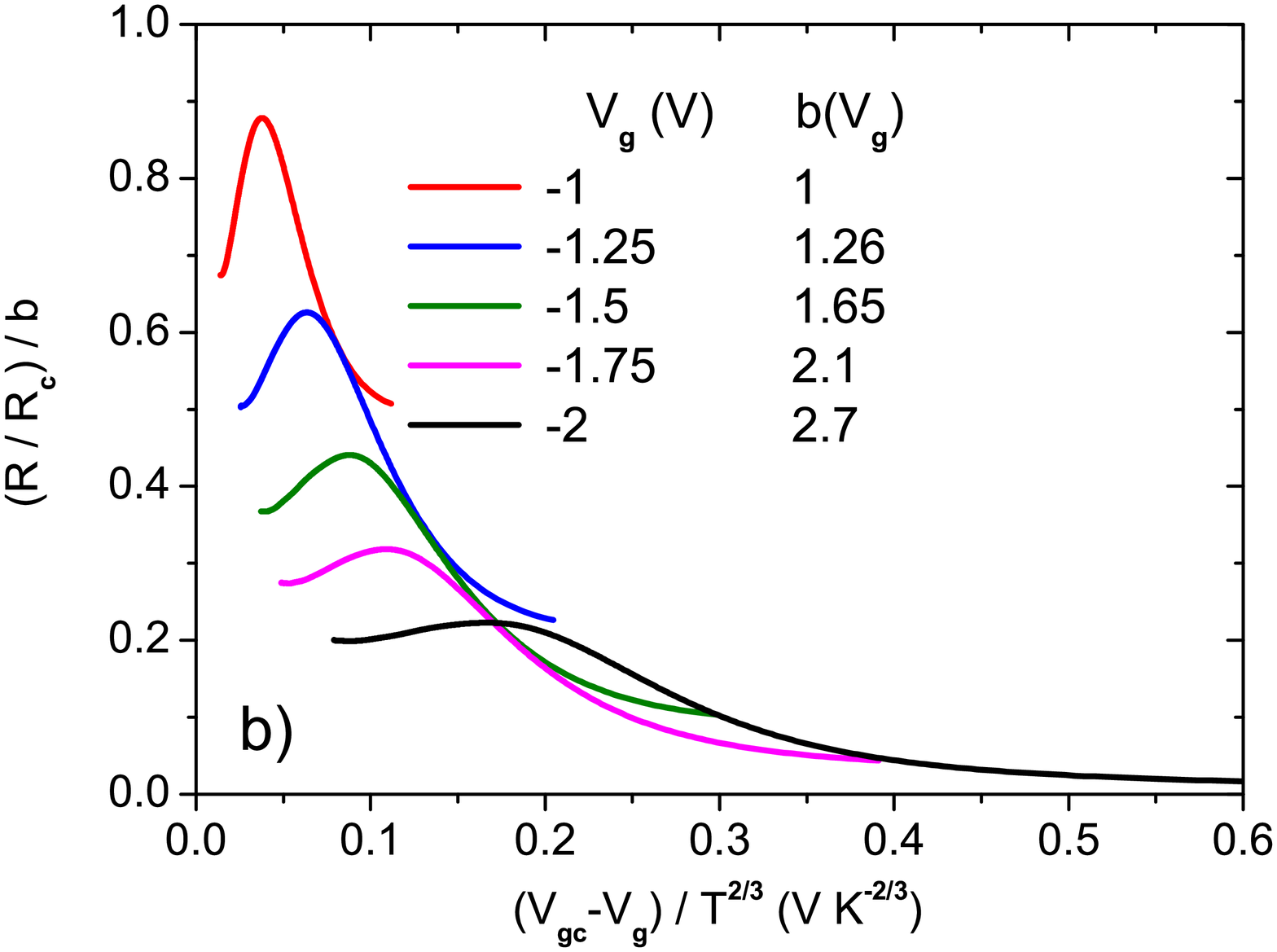} \vspace{-0.5cm}
\caption{(color online) a) $R/R_{c}$ \textit{vs}.$\left( V_{gc}-V_{g}\right)
/T^{2/3}$ with $R_{c}=6.45$ k$\Omega $ and $V_{gc}=-0.7$ V derived from the
temperature dependence of the sheet resistance of Bollinger \textit{et al}.%
\protect\cite{bol} The data cover the range from $4$ K to $100$ K. b) $%
R/b\left( V_{g}\right) R_{c}$ vs.$\left( V_{gc}-V_{g}\right) /T^{2/3}$ with $%
b\left( V_{g}=-1 V\right) =1$ and $b\left( V_{g}\right) $ adjusted to
achieve a piecewise collapse of the data.}
\label{fig5}
\end{figure}

\begin{figure}[tbh]
\includegraphics[width=1.0\linewidth]{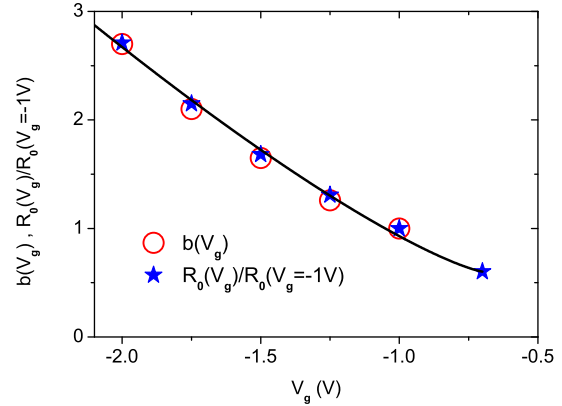} \vspace{-0.5cm}
\caption{(color online) $b\left( V_{g}\right) $ and $%
R_{0}(V_{g})/R_{0}(V_{g}=-1$V$)$. The solid line is $%
R_{0}(V_{g})/R_{0}(V_{g}=-1$V$)=\left( R_{c}+15.9\left( V_{gc}-V_{g}\right)
^{1.26}\right) /R_{0}(V_{g}=-1$V$)$ with $R_{c}=6.45$ k$\Omega $, $%
V_{gc}=-0.7$ and $R_{0}(V_{g}=-1$V$)$ =10.7 k$\Omega $ taken from Fig.
\protect\ref{fig2}b.}
\label{fig6}
\end{figure}

To complete the analysis of the data of Bollinger \textit{et al}.\cite{bol}
we depicted in Fig. \ref{fig4} the plot $R\left( V_{g},T\right) /R_{0}\left(
V_{g}\right) $ \textit{vs}.$\left( T_{c}\left( V_{g}\right) /T\right) ^{2/3}$%
corresponding to the quantum scaling function $G\left( y\right) $ (Eq. (\ref%
{eq7})) in terms of $y/y_{c}=\left( T_{c}\left( V_{g}\right) /T\right)
^{2/3} $. Apparently, the data does not fall completely on the BKT curve
indicated by the dashed line. It corresponds to the superconductor branch of
the quantum scaling function. Instead we observe a flow to and away from the
universal characteristics. As $T_{c}/T$ decreases for fixed $T_{c}$ the
crossover to AL-behavior sets in, while the rounding of the transition leads
with increasing $T_{c}/T$ to a flow away from criticality. The important
lesson then is that the quality of the data collapse on a single curve
heavily depends on the temperature range of the data entering the plot.
Another striking feature is the extended scaling regime. Within the BKT
scenario it simply follows from the fact that the scaling form (\ref{eq3})
applies along the BKT transition line irrespective of the distance from the
QSI transition. In the quantum scaling approach this property remains hidden
and merely suggests an extended quantum critical regime. The excellent
quality of the piecewise data collapse also reveals that the observance of
the substantial variation of $R_{0}\left( V_{g}\right) $ (see Fig. \ref{fig2}%
b) is essential, while in the quantum scaling approach it is fixed by the
critical sheet resistance.
To demonstrate these features even more compelling we depicted in Fig. \ref%
{fig5}a $R/R_{c}$ \textit{vs}.$\left( V_{gc}-V_{g}\right) /T^{2/3}$.
Apparently the data do not fall even piecewise on a single curve. After all
this is not surprising because the quantum scaling form (\ref{eq1}) holds
close to the critical sheet resistance only and $R_{0}\left( V_{g}\right) $
varies substantially in the gate voltage regime considered here (see Fig. %
\ref{fig2}b). As shown in Fig. \ref{fig5}b and Fig. \ref{fig6} this behavior
allows to estimate $R_{0}\left( V_{g}\right) $ from $R/R_{c}$ \textit{vs}.$%
\left( V_{gc}-V_{g}\right) /T^{2/3}$ by rescaling $R_{c}$ in terms of $%
b\left( V_{g}\right) R_{c}$ with $b\left( V_{g}=-1V\right) =1$ to achieve
piecewise a collapse of the data. The resulting $b\left( V_{g}\right) $ is
shown in Fig. \ref{fig6} and agrees well with $R_{0}(V_{g})/R_{0}(V_{g}=-1$V$%
)$ derived from the BKT behavior of the sheet resistance (see Fig. \ref{fig2}%
b).

A BKT line with a QSI transition at its endpoint was also explored rather
detailed at the interface between the insulating oxides LaAlO$_{3}$ and SrTiO%
$_{3}$ exhibiting a superconducting 2D electron system that can be modulated
by a gate voltage.\cite{tsintf,reyren,cavigli,bert} BKT behavior and with
that a 2D electron system was established as follows: The the
current-voltage characteristics\cite{reyren} revealed at the BKT transition
temperature $T_{c}$ the characteristic BKT form $V\propto I^{a}$ with $a=3$.
\cite{halp} Consistency with the characteristic temperature dependence of
the sheet resistance (Eq. (\ref{eq3})) was established.\cite%
{tsintf,reyren,cavigli} It was also shown that the effective thickness of
the superconducting 2D system can be extracted from the magnetic field
dependence of the conductivity at $T_{c}$.\cite{tstool} The gate voltage
tuned BKT phase transition line, $T_{c}\left( V_{g}\right) $, derived from
the temperature dependence of the sheet resistance at various gate voltages
uncovered with Eq. (\ref{eq3}) consistency with $T_{c}\left( V_{g}\right)
=8.9\times 10^{-3}$ $\left( V_{g}-V_{gc}\right) ^{2/3}$ K \ pointing to
quantum critical behavior (Eq. (\ref{eq2})) with $z\overline{\nu }=2/3$ and
the critical sheet resistance $R_{c}=2.7$ k$\Omega $.\cite{tsintf,cavigli}
Furthermore the estimates $\overline{\nu }\simeq 2/3$ and $z\simeq 1$,
confirming $z\overline{\nu }=2/3$, have been derived from $R_{0}\left(
V_{g}\right) $. \cite{tsintf} This suggests that the gate voltage tuned QSI
transition of the 2D electron system at the LaAlO$_{3}$/ SrTiO$_{3}$
interface belongs to the $(2+z)=3$-xy universality class where hyperscaling
and with that the proportionality between $d/\lambda ^{2}\left( 0\right) $
and $T_{c}$ (Eq. (\ref{eq10})) applies. On the other hand because $\overline{%
\nu }\simeq 2/3<2/D=1$ disorder is according to Harris theorem relevant as
well.\cite{harris,chay}

\begin{figure}[tbh]
\includegraphics[width=1.0\linewidth]{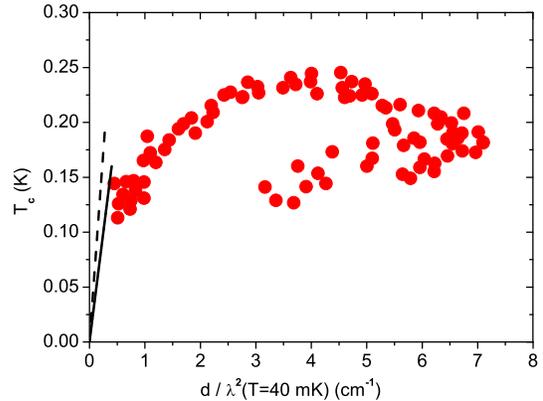} \vspace{-0.5cm}
\caption{(color online) $T_{c}$ \textit{vs}. $d/\protect\lambda ^{2}\left(
T=0.04K\right) $ for a gate voltage tuned superconducting LaAlO$_{3}$/SrTiO$%
_{3}$ interface taken from Bert \emph{et al}.\protect\cite{bert} The solid
and dashed lines are the predicted universal relationship (\protect\ref{eq8}%
) close to the QSI transition with $Q_{2}=2/\protect\pi $, resulting from
the lower bound (\protect\ref{eq11}), and $Q_{2}=1.12$ derived from the $%
^{4} $He film data of Crowell \textit{et al}.\protect\cite{crow}}
\label{fig7}
\end{figure}

To comment the difficulties in observing the BKT features in the magnetic
penetration depth as well as the relation between $d/\lambda ^{2}\left(
0\right) $ and $T_{c}$, we reproduced in Fig. \ref{fig7} the data of Bert
\textit{et al}.\cite{bert} for a gate voltage tuned superconducting LaAlO$%
_{3}$/SrTiO$_{3}$ interface in terms of $T_{c}$ \textit{vs}. $d/\lambda
^{2}\left( T=0.04K\right) $. $T_{c}$ is defined here as the temperature at
which the diamagnetic screening drops below the noise level corresponding to
a detectable $d/\lambda ^{2}$ of $0.10-0.34$ cm$^{-1}$. A glance at Fig. \ref%
{fig7} reveals that this is just the regime where the universal quantum
behavior applies, indicated by the dashed and solid lines, corresponding to
the lower bound (\ref{eq11}) and the behavior derived from the $^{4}$He data
of Crowell \textit{et al}.,\cite{crow} respectively. Nevertheless the data
reveal the flow to the 2D-QSI transition which is attained at much lower $%
T_{c}$'s. Otherwise the data points resemble the outline of a fly's wing,%
\cite{parks} remarkably similar to the $T_{c}$ \textit{vs.} $1/\lambda
_{ab}^{2}\left( T=0\right) $ plots of the bulk superconductors Y$_{0.8}$Cu$%
_{0.2}$-123, Tl-1212,\cite{bern} and Tl-2201\cite{nied}, covering nearly the
doping regime of the so called superconducting dome extending from the
underdoped to the overdoped limit . According to the generic plot $%
T_{c}/T_{c}\left( x_{m}\right) $ \emph{vs}. $\gamma \left( x_{m}\right)
/\gamma \left( T_{c}\right) $, where $\gamma =\xi _{ab}/\xi _{c}$ is the
anisotropy and $\xi _{ab,c}$ denote the in-plane and c-axis correlation
length, these cuprates become nearly 2D in the underdoped limit.\cite{parks}
In any case, Fig. \ref{fig7} shows that in this plot the universal QSI
behavior is attained at comparatively low $T_{c}$ values only. Accordingly,
in the QSI regime of interest the Nelson-Kosterlitz jump given by Eq. (\ref%
{eq9}) becomes very small and appears to be beyond present experimental
resolution.\cite{bert} On the contrary in the temperature dependence of the
sheet resistance is the BKT critical regime accessible because $%
T_{c}/T_{c0}\simeq 0.75$,\cite{tstool} even though small compared to that in
the La$_{2-x}$Sr$_{x}$CuO$_{4}$ films where $T_{c}/T_{c0}\approx 0.1$.

\section{Summary and discussion}

In sum, we sketched and explored the implications of
Berezinskii-Kosterlitz-Thouless (BKT) critical behavior on the quantum
critical properties of a two dimensional (2D) quantum
superconductor-insulator transition (QSI) driven by the tuning parameter $x$%
. It was shown that the finite temperature BKT scenario, implies in terms of
the characteristic temperature dependence of the BKT correlation length an
explicit quantum scaling function for the sheet resistance $R\left(
x,T\right) $ along the superconducting branch ending at the nonuniversal
critical value $R_{c}=R_{0}\left( x_{c}\right) $. This scaling form fixes
the BKT-transition line $T_{c}\left( x\right) $ and provides estimates for
the quantum critical exponent product $z\overline{\nu }$. In addition,
independent estimates of $z\overline{\nu }$, $z$ and $\overline{\nu }$
follow from the $x$ dependence of the nonuniversal parameters $T_{c}\left(
x\right) $ and $R_{0}\left( x\right) $ entering the characteristic BKT
expression for the sheet resistance $R\left( x,T\right) $. This requires
that the BKT critical regime where phase fluctuations dominate is attained
and the finite temperature BKT relation for the sheet resistance applies for
any $T\geq T_{c}\left( x\right) >0$. The last condition is satisfied because
the BKT expression for the sheet resistance is simply related to the
characteristic temperature dependence of the BKT correlation length. Quantum
fluctuations enter via the nonuniversal parameters $T_{c}\left( x\right) $
and $R_{0}\left( x\right) $ disclosing close to $x_{c}$ the respective
quantum critical behavior. Even though BKT critical behavior is not affected
by short-range correlated and uncorrelated disorder\cite{harris,chay} the
observation of this requires sufficiently homogeneous films and a tuning
parameter which does not affect the disorder. Noting that in the magnetic
field tuned case there is no BKT-line, thickness and gate voltage tuned
2D-QSI transitions appear to be promising candidates. In any case the
scenario outlined here requires a line of BKT-transitions $T_{c}\left(
x\right) $ with a quantum critical endpoint $T_{c}\left( x_{c}\right) =0$
and sheet resistance data which attain the BKT critical regime.

To illustrate the potential and the implications of this scenario we
analyzed data of Bollinger \textit{et al}.\cite{bol} taken on gate voltage
tuned epitaxial films of La$_{2-x}$Sr$_{x}$CuO$_{4}$ that are one unit cell
thick. Evidence for dominant phase fluctuations and BKT-critical behavior
was established in terms of the temperature dependence of the sheet
resistance revealing a large critical regime extending substantially above
the lowest attained temperature $T=4$ K. From the nonuniversal parameters $%
T_{c}\left( x\right) $ and $R_{0}\left( x\right) $ disclosing the respective
quantum critical properties we derived for the quantum critical exponents
the estimates: $z\overline{\nu }\simeq 1.46$ from $T_{c}\left( x\right) $, $%
\overline{\nu }\simeq 0.63$ and $z\simeq 2.35$ from $R_{0}\left( x\right) $,
yielding $z\overline{\nu }\simeq 1.48$. Thus, in contrast to the standard
quantum scaling approach, providing an estimate for $z\overline{\nu }$ only,
the BKT scenario uncovered $z\overline{\nu }$, $\overline{\nu }$ and $z$
from the quantum critical behavior disclosed in $T_{c}\left( x\right) $ and $%
R_{0}\left( x\right) $. Additional evidence for $z\overline{\nu }\simeq 1.5$
was established from the comparison of the scaled data with the explicit
scaling BKT scaling form of the superconductor branch. We observed that the
scaled data does not fall entirely on the BKT curve. Instead a flow to and
away from the universal characteristics occurred. As $T_{c}/T$ decreases for
fixed $T_{c}$ a crossover to AL-behavior sets in, while the rounding of the
transition leads with increasing $T_{c}/T$ to a flow away from criticality.
The important lesson then is that the quality of the data collapse on a
single curve heavily depends on the temperature range of the data entering
the plot. Another striking feature is the extended scaling regime. Within
the BKT scenario it follows from the fact that the explicit scaling form of
the sheet resistance applies along the entire BKT transition line
irrespective of the distance from the QSI transition. In the quantum scaling
approach this property remains hidden and merely suggests an extended
quantum critical regime. The piecewise excellent quality of the data
collapse also reveals that the provision of the substantial variation of $%
R_{0}\left( x\right) $ is essential, while in the quantum scaling approach
it is fixed by the critical sheet resistance. Supposing that the equivalence
between quantum phase transitions with $D$ spatial dimensions and the ones
of classical phase transitions in $\left( D+z\right) $ dimensions applies,
the 2D-QSI transition at the endpoint of a BKT line $T_{c}\left( x\right) $
should belong to the finite temperature $\left( 2+z\right) -xy$ universality
class. Our estimate $z\simeq 2.35$ and with that $D=4.35-xy$ critical
behavior where $\overline{\nu }=1/2$. However the fate of this clean
critical point under the influence of disorder is controlled by the Harris
criterion.\cite{harris,chay} If the inequality \ $\overline{\nu }\geq 2/D$
is fulfilled, the disorder does not affect the critical behavior. If the
Harris criterion is violated (\ $\overline{\nu }<2/D$), the generic result
is a new critical point with conventional power law scaling but new
exponents which fulfill \ $\overline{\nu }<2/D$. Since $\overline{\nu }=1/2$
violates this inequality disorder is relevant and drives the system from the
mean-field to an other critical point with different critical exponents as
our estimate, $\overline{\nu }\simeq 0.63$, consistent with $\overline{\nu }%
<2/D=1$, uncovers. The resulting 2D-QSI transition with $\overline{\nu }%
\simeq 0.63$ and $z\simeq 2.35$ violates then the equivalence between
quantum phase transitions in systems with $D$ spatial dimensions and the
ones of classical phase transitions in $\left( D+z\right) $ dimensions. In
addition the proportionality between $d/\lambda ^{2}\left( 0\right) $ and $%
T_{c}$ (Eq. (\ref{eq10})), valid below the upper critical dimension $D_{u}=4$%
, does no longer hold because $\left( D+z\right) \simeq 4.35$ is above $%
D_{u}=4$. In fact magnetic penetration depth measurements taken on
underdoped high quality YBa$_{2}$Cu$_{3}$O$_{6+x}$ single crystals revealed $%
T_{c}\propto \left( d/\lambda ^{2}\left( 0\right) \right) ^{0.61}$.\cite%
{liang} Another option is that the disorder destroys the QSI transition.
Given the evidence for a rounded BKT transition (see Fig. \ref{fig1}) and
noting that the analyzed data do not extend very close to the QSI transition
(see Fig. \ref{fig4}) further studies are required to elucidate this option.
In any case, our estimate $\overline{\nu }\simeq 0.63$, the evidence for BKT
behavior at finite temperature and the Harris theorem imply the presence and
relevance of disorder at the QSI transition and its irrelevance at finite
temperature. Further evidence for the importance of disorder follows from
the temperature dependence of the sheet resistance in the insulating phase.
Considering $R\left( Vg=0,T\right) $ of Bollinger \textit{et al}.\cite{bol}
we observe consistency with the Mott variable range hopping model in 2D. The
conductivity exhibits the characteristic temperature dependence $\sigma
=\sigma _{0}\exp \left( -\left( T_{0}/T\right) ^{1/\left( D+1\right)
}\right) $ which applies to strongly disordered systems with localized
states.\cite{mott}

A previous and analogous analysis of the sheet resistance data of the
superconducting LaAlO$_{3}$/ SrTiO$_{3}$ interface revealed the critical
sheet resistance $R_{c}=2.7$ k$\Omega $ and the exponents $z\overline{\nu }%
\simeq 2/3$, $\overline{\nu }\simeq 2/3$ and $z\simeq 1$.\cite{tsintf} In
this case the gate voltage tuned QSI transition of the 2D electron system at
the LaAlO$_{3}$/ SrTiO$_{3}$ interface has a finite temperature counterpart
in $\left( D+z\right) $ dimensions namely the $(2+z)=3$-xy model where
hyperscaling and with that the proportionality between $d/\lambda ^{2}\left(
0\right) $ and $T_{c}$ (Eq. (\ref{eq10})) applies. On the other hand $%
\overline{\nu }\simeq 2/3<2/D=1$ implies according to Harris theorem\cite%
{harris,chay} that disorder is relevant as well.

To comment on the BKT-features in the temperature dependence of the magnetic
penetration depth we considered the data of Bert \textit{et al}.\cite{bert}
for a gate voltage tuned superconducting LaAlO$_{3}$/SrTiO$_{3}$ interface
in terms of $T_{c}$ \textit{vs}. $d/\lambda ^{2}\left( T=0.04\text{ K}%
\right) $. The data reveals the flow to the universal relationship (\ref%
{eq10}) but much lower $T_{c}$ must be attained to reach quantum critical
regime. However in this low$T_{c}$ regime is the Nelson-Kosterlitz jump
given by Eq. (\ref{eq9}) very small and appears to be beyond present
experimental resolution.\cite{bert} Otherwise the data points resemble the
outline of a fly's wing,\cite{parks} remarkably similar to the $T_{c}$
\textit{vs.} $1/\lambda _{ab}^{2}\left( T=0\right) $ plots of the bulk
superconductors Y$_{0.8}$Cu$_{0.2}$-123, Tl-1212,\cite{bern} and Tl-2201\cite%
{nied}, covering nearly the entire doping regime in the so called
superconducting dome extending from the underdoped to the overdoped limit.

\end{document}